\documentclass[journal]{IEEEtran}
\usepackage[dvipdfmx]{graphicx,xcolor}
\usepackage{epsfig}
\usepackage[T1]{fontenc}
\usepackage{lmodern}
\usepackage{textcomp}
\usepackage{latexsym}
\usepackage{cite}
\usepackage{bm}
\usepackage{amsmath}
\usepackage{amsfonts}

\newcommand{\ee}        {{\rm e}}
\newcommand{\vct}[1]    {\mbox{\boldmath{$#1$}}}
\newcommand{\jj}        {{\rm j}}
\newcommand{\dd}        {{\rm d}}

\newcommand{\HH}        {{\bf H}}
\newcommand{\TT}        {\mathrm{T}}

\newcommand{\astr}      {^{\rm *}}        

\begin{document}
\title{Multiradar Data Fusion for Respiratory Measurement of Multiple People}
\author{Shunsuke Iwata, \IEEEmembership{Student Member, IEEE,} Takato~Koda, and Takuya~Sakamoto,~\IEEEmembership{Senior Member,~IEEE,}
  \thanks{This study was supported in part by JSPS KAKENHI under Grants 19H02155 and 21H03427, JST PRESTO under Grants JPMJPR1873, and JST COI under Grants JPMJCE1307.}
  \thanks{S.~Iwata, T.~Koda and T.~Sakamoto are with the Department of Electrical Engineering, Graduate School of Engineering, Kyoto University, Kyoto, Kyoto, 615-8510 Japan.}}
\markboth{}%
{Iwata \emph{et al.}: Multiradar Data Fusion for Respiratory Measurement of Multiple People}

\maketitle

\begin{abstract}
This study proposes a data fusion method for multiradar systems to enable measurement of the respiration of multiple people located at arbitrary positions. Using the proposed method, the individual respiration rates of multiple people can be measured, even when echoes from some of these people cannot be received by one of the radar systems because of shadowing. In addition, the proposed method does not require information about the positions and orientations of the radar systems used because the method can estimate the layout of these radar systems by identifying multiple human targets that can be measured from different angles using multiple radar systems. When a single target person can be measured using multiple radar systems simultaneously, the proposed method selects an accurate signal from among the multiple signals based on the spectral characteristics. To verify the effectiveness of the proposed method, we performed experiments based on two scenarios with different layouts that involved seven participants and two radar systems. Through these experiments, the proposed method was demonstrated to be capable of measuring the respiration of all seven people by overcoming the shadowing issue. In the two scenarios, the average errors of the proposed method in estimating the respiration rates were 0.33 and 1.24 respirations per minute (rpm), respectively, thus demonstrating accurate and simultaneous respiratory measurements of multiple people using the multiradar system.
\end{abstract}
\IEEEpeerreviewmaketitle

\section{Introduction}
\IEEEPARstart{R}{espiratory} measurements have become increasingly important in a variety of applications, including healthcare, home medication, medical alert systems for babies, and monitoring systems used for infants and senior citizens. A variety of sensors and systems have been developed for these applications to monitor patient respiration \cite{review1,review2,review3} and among these systems, radar-based systems have been attracting attention as a promising approach because they are suitable for performing long-term measurements without patient discomfort since there is no need for the patient to wear a device. A number of existing studies have been reported on the measurement of people using radar techniques, including systems for estimation of the locations of people \cite{location1,location2,location3,location4,location5,location6}, tracking of people in motion \cite{tracking1,tracking2,tracking3,tracking4,tracking5,tracking6,tracking7,tracking8}, and detection of patient vital signs, including their heartbeat and respiration \cite{respiration1,respiration2,respiration3,respiration4,respiration5,respiration6,respiration7,respiration8,respiration9,respiration10,respiration11,respiration12,respiration13,respiration14,respiration15,respiration16,respiration17,respiration18,respiration19,respiration20,respiration21}.

Another advantage of radar-based respiratory measurement is the possibility of performing simultaneous measurements of the respiration of multiple people using only a single device. There have been numerous studies in which a single radar system has been used to measure the respiration of multiple people \cite{respiration1,respiration16,respiration17,respiration18,respiration19,respiration20,respiration2,respiration3,respiration4,respiration5,respiration6,respiration7,respiration8,respiration9,respiration12,respiration10,respiration11,respiration13,respiration21}. For example, Nosrati et al. \cite{respiration10}, Islam et al. \cite{respiration11}, Su et al. \cite{respiration13}, and Koda et al. \cite{respiration21} used single radar systems to demonstrate simultaneous measurement of the respiration of two, three, three, and seven people, respectively. These studies, however, did not consider the effect of the shadowing problem, in which an echo from one target person is blocked by another person located between the first person and the radar system; this results in the radar system being unable to detect a person (or some people) when multiple people are present in the scene.

To overcome the shadowing problem, this study introduces simultaneous use of multiple radar systems to perform respiratory measurements of multiple people. There have been previous studies that used multiple radar systems to measure the respiration of multiple people: Shang et al. \cite{respiration15} used two radar systems to measure two people in the presence of body motion, and Yang et al. \cite{respiration14} also used two radar systems to measure five people. In these studies, it was assumed that the relative positions and orientations of the radar systems were known in advance, and that the echoes from all target people could be measured directly by each of the radar systems without the shadowing problem. These assumptions, however, cannot always be satisfied in practice, which hinders the use of the radar-based respiratory measurement approach in actual daily environments.

As discussed above, none of the existing studies have accounted for the effect of the shadowing problem, which prevents the echo from one person from being detected by one of the radar systems. In this study, we propose a novel multiradar data fusion method to perform respiratory measurements that can measure multiple people simultaneously. The proposed method can mitigate the effect of the shadowing problem using multiple radar systems without knowing the relative locations and orientations of these radar systems. The proposed method generates multiple radar images using the multiradar data, and these radar images are then converted into multiple sets of two-dimensional point clouds. To align these point clouds, the proposed method then estimates the relative positions and orientations of the multiple radar systems, which allows for data fusion of the multiradar system and thus overcomes the shadowing problem. The proposed method even works when the positional relationships among the multiple radar systems are unknown, which means that there is no requirement for calibration to be performed beforehand, and also no restriction on the physical installation of the radar systems. The performance of the proposed method is evaluated quantitatively by performing radar measurements involving seven participants and a pair of radar systems.

\section{Imaging and Clustering of Human Targets Using Multiple Radar Systems}
We assume use of a multiradar system comprising $M$ radar systems with positions and angles that are unknown. Each radar system contains a $K$-element array (or virtual array). Let $s_i(t,r)$ denote the signal received by the $i$-th element $(i=0,1,\cdots,K-1)$ of the array, where $t$ is the slow time and the range $r$ is expressed as $r=ct'/2$ using the fast time $t'$ and the speed of light $c$. A signal vector $\vct{s}(t,r)$ is defined as $\vct{s}(t,r) = [s_0(t,r), s_1(t,r), \cdots , s_{K-1}(t,r)]^\TT$, where the superscripted T represents a transpose operator. We assume here that all elements of the array antennas are calibrated and that static clutter is suppressed by subtracting the time-averaged signal. 

\begin{figure}[bt]
    \begin{center}
      \includegraphics[width=0.6\linewidth]{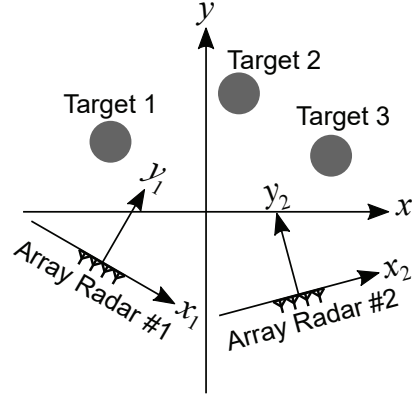}
      \caption{Schematic of system model for measurement of multiple target people using multiple radar systems.}
        \label{fig1}
    \end{center}
\end{figure}

Using the $m$-th radar system ($m=1,\cdots,M$), a radar image is generated and is expressed in the two-dimensional Cartesian coordinate system $(x_m, y_m)$, where the $x_m$ axis is aligned with the array baseline of the $m$-th radar system. A complex radar image $I_m(t,\vct{r}_m)$ is generated from the $m$-th radar system's data as $I_m(t,\vct{r}_m)=\vct{w}(\phi_m)^\HH \vct{s}(t,\rho_m)$, where $\vct{r}_m$ is the position vector that can be expressed as $(x_m, y_m)$ in Cartesian coordinates and as $(\rho_m,\phi_m)$ in polar coordinates. The weight vector is $\vct{w}(\theta)=[w_0,w_1,\cdots,w_{K-1}]^\mathrm{T}$, where the choice of $w_i$ depends on the imaging algorithm being used. In this study, we use a straightforward beamformer (BF) method for simplicity, in which the weights are set as $w_i(\theta)=\alpha_i\ee^{-\jj (2\pi x_i/\lambda)\sin\theta}$ $(i=0,1,\cdots,K-1)$. Here, we use the Taylor window coefficient $\alpha_i$. Fig.~\ref{fig1} shows a system model containing multiple radar systems for measurement of multiple human targets, in which we see that each radar system has its own coordinate system.

Next, a point cloud is generated from each radar image; the $m$-th point cloud is generated randomly to follow a density distribution that is proportional to $\rho_m|I_m(t,\vct{r}_m)|^2$, where $\rho_m$ is introduced to compensate for propagation losses. The X-means clustering technique \cite{respiration21,xmeans} is then applied to the point cloud and the clusters $c_1, c_2, \cdots, c_N$ are generated; these clusters correspond to the $N$ targets. For presentation purposes, we generated a radar cluster image $C_m(t,\vct{r}_m)$ that takes the values $\{0,1,2,\cdots,N\}$, which correspond to the indices of the clusters $c_1, c_2, \cdots, c_N$ except for 0; $C_m(t,\vct{r}_m)=0$ indicates that no cluster is associated with $\vct{r}_m$. Finally, the representative position $\vct{r}_m^{(i)}$ of cluster $c_i$ in the $m$-th radar image is calculated \cite{respiration21}; the clusters' representative positions are then processed in the proposed data fusion algorithm, which combines multiple radar images.

\section{Multiradar Data Fusion for Respiratory Measurement of Multiple People}
The main issue when performing radar measurements of multiple human targets indoors is that not all of these human targets can be measured directly using only a single radar system because of shadowing, i.e., where one human echo is blocked by another human target or by obstacles; this occurs frequently when multiple people are located together and are densely spaced. To mitigate this issue, we use a multiradar system to increase the probability that at least one of these radar systems can detect all human targets in the scene. Given the radar images $I_m(t,\vct{r}_m)$ and the cluster positions $\vct{r}_m^{(i)}$ for $m=1,\cdots, M$, we propose a data fusion method here that combines the multiple radar images. Here, we assume a special case with $M=2$ for simplicity. The first step in the method is to find a rigid transformation that can align the coordinates from the $x_2$-$y_2$ coordinate system to those of the $x_1$-$y_1$ coordinate system.

\subsection{Procrustes Analysis for Alignment of Associated Target Clusters}
\label{subsec:procrustes}
We propose a data fusion method that uses Procrustes analysis \cite{procrustes} to find the rigid transformation parameters required to describe the relationship between two sets of points in a plane. Let us assume here that $N$ pairs of two-dimensional vectors $(\vct{r}_1^{(1)},\vct{r}_2^{(1)}), (\vct{r}_1^{(2)},\vct{r}_2^{(2)}), \cdots,(\vct{r}_1^{(N)},\vct{r}_2^{(N)})$ correspond to $N$ pairs of points. The purpose of this section is to find a rigid transformation that adjusts the points $\vct{r}_2^{(n)}$ ($n=1,\cdots,N$) with respect to $\vct{r}_1^{(n)}$ ($n=1,\cdots,N$). If appropriate transformation parameters are selected, the points are then transformed to satisfy $\vct{r}_1^{(n)} \simeq R\vct{r}_2^{(n)}+\vct{t}$ using the translation vector $\vct{t}$ and the rotation matrix $R$. If we subtract the average coordinates $\bar{\vct{r}}_m = (1/N)\sum_{n=1}^N\vct{r}_m^{(n)}$ ($m=1,2$) from the position vectors using $\tilde{\vct{r}}_m^{(n)} = \vct{r}_{m}^{(n)}-\bar{\vct{r}}_m$, we can then simplify the problem for large $N$; we need only estimate the $R$ that satisfies $\tilde{\vct{r}}_1^{(n)}=R\tilde{\vct{r}}_2^{(n)}$. Here, we define $2\times N$ matrices $S_1$ and $S_2$ as:
\begin{align}
  S_1&=[\tilde{\vct{r}}_{1}^{(1)},\tilde{\vct{r}}_{1}^{(2)}, \cdots, \tilde{\vct{r}}_{1}^{(N)}],\\
  S_2&=[\tilde{\vct{r}}_{2}^{(1)},\tilde{\vct{r}}_{2}^{(2)}, \cdots, \tilde{\vct{r}}_{2}^{(N)}],
\end{align}
and we estimate a rotation matrix $R\astr$ as:
\begin{equation}
  \begin{array}{ccl}
    R^\ast&=&\arg\min_{R}\|S_1-RS_2\|^{2}_{\mathrm{F}},\\
    &&\mathrm{subject\ to}\ R^{\mathrm{T}}R=I, \det R = 1,
  \end{array}
  \label{procrustes2}
\end{equation}
where the subscripted F denotes the Frobenius norm and represents the square root of the sum of squares of all components.

In the Procrustes analysis, rather than solve for the optimization problem in Eq.~(\ref{procrustes2}) directly, singular value decomposition is used to estimate $R^\ast$; $S_1S_2^\TT$ is decomposed using the form $S_1S_2^\TT = U\Sigma V^\TT$, where $U$ and $V$ are $2\times 2$ orthogonal matrices, and $\Sigma$ represents a $2\times 2$ diagonal matrix with non-negative real numbers. We then obtain the rotation matrix $R^\ast=UV^\TT$. Using $R^\ast$, the position vector $\tilde{\vct{r}}_2^{(n)}$ is transformed to give $R^\ast \tilde{\vct{r}}_2^{(n)}+\bar{\vct{r}}_1$, which is a mapping from the coordinate system $x_2$-$y_2$ to the coordinate system $x_1$-$y_1$. Although we have discussed the special case where $M=2$ above, a similar procedure can be performed in a general case, including cases where $M>2$.

\subsection{Proposed Data Fusion Method Using Respiratory Correlation and Procrustes Analysis}
To apply Procrustes analysis to transformation of the coordinate systems of a multiradar system, the cluster positions can be used to form the input matrices $S_1$ and $S_2$. We should note here that Procrustes analysis requires at least two pairs of corresponding points, e.g., $(\vct{r}_1^{(1)},\vct{r}_2^{(1)}), (\vct{r}_1^{(2)},\vct{r}_2^{(2)})$. In this section, we propose a method to associate the cluster positions that uses the respiratory features of the human targets. In the proposed method, the similarity of the respiratory displacement waveforms is used to associate multiple echoes. Although we explain the procedure for the case where $M=2$ for simplicity, the procedure can be extended to a general case with $M>2$. The proposed method comprises the following five steps.
\begin{enumerate}
\item Two pairs of corresponding points are estimated using the correlation coefficient for the respiratory displacement waveforms. For the position vectors $\vct{r}_1^{(n_1)}$ and $\vct{r}_2^{(n_2)}$, the correlation coefficient is calculated as:
  \begin{equation}
    \rho_\mathrm{c}({n_1,n_2}) = (1/\rho_0)\int \angle I_1(t,\vct{r}_1^{(n_1)}) \angle I_2(t,\vct{r}_2^{(n_2)}) \dd t,
  \end{equation}
  where $\rho_0$ is introduced for normalization. First, we find a pair $(n_1\astr,n_2\astr)$ that maximizes $\rho_\mathrm{c}(n1,n2)$, where
  \begin{equation}
    (n_1\astr,n_2\astr)=\arg\max_{(n_1,n_2)}\rho_\mathrm{c}(n1,n2).
  \end{equation}
  Then, we find another pair $(n_1^{**},n_2^{**})$, where
  \begin{equation}
    \begin{array}{ccc}
      (n_1^{**},n_2^{**})&=&\arg\max_{(n_1,n_2)}\rho_\mathrm{c}(n1,n2)\\
      &&\mathrm{subject\ to}\ n_1\neq n_1\astr, n_2\neq n_2\astr.
    \end{array}
  \end{equation}
\item Procrustes analysis is applied to the two pairs of points $(n_1\astr,n_2\astr)$ and $(n_1^{**},n_2^{**})$, and the rigid transformation parameters $R\astr$ and $\vct{t}\astr$ are then obtained.
\item Using $R\astr$ and $\vct{t}\ast$, all position vectors $\vct{r}_m^{n}$ $m=1,\cdots,M; n=1,\cdots,N$ are transformed into the coordinate system $x_1$-$y_1$. All pairs of transformed points separated by a distance of less than a threshold distance (i.e., $|\vct{r}_1^{(n_1)}-R\astr\vct{r}_2^{(n_2)}-\vct{t}\astr| \leq d_\mathrm{th}$) are regarded as associated pairs.\item All associated pairs are processed using the Procrustes analysis, and the transformation parameters $R\astr$ and $\vct{t}\astr$ are then updated. 
\item For each human target, one of the displacements $\angle I_1(t,\vct{r}_1^{(n_1)}), \cdots, \angle I_M(t,\vct{r}_1^{(n_M)})$ is selected that is most likely to reflect the target's respiratory motion. We proposed selection of the displacement with the waveform that is closest to a sinusoidal wave. Using the following fourth-order moment of the power spectrum, we then find the $m$ that maximizes $\kappa(n,m)$ for each $n$ ($1\leq n\leq N$). We define
\begin{equation}
  \kappa(n,m)=\frac{\int^\infty_0 \left|\int^\infty_{-\infty}I_m(t,\vct{r}_m^{(n)})\mathrm{e}^{-\jj\omega t}\mathrm{d}t\right|^4\mathrm{d}\omega}{\left|\int^\infty_0\left|\int^\infty_{-\infty}I_m(t,\vct{r}_m^{(n)})\mathrm{e}^{-\jj\omega t}\mathrm{d}t\right|^2\mathrm{d}\omega\right|^2},
\end{equation}
and the optimum $m$ is then found as
\begin{equation}
  m\astr(n) = \arg\max_{m}\kappa(n,m),
\end{equation}
which means that the $m\astr(n)$-th radar signal is used for the $n$-th human target. As a result, the displacement waveform that is most likely to be related to their respiration is selected for each person, thus leading to accurate measurement of their respiration.
\end{enumerate}

\section{Experimental Evaluation of the Proposed Method}
\subsection{Multiradar Experiment with Multiple People}
We performed the experiments using both a pair of array radar systems and belt-type respirometers simultaneously to evaluate the accuracy of the multiradar-based respiratory measurements acquired using the proposed method. To evaluate the accuracy of the method in measuring the respiratory intervals, all participants wore a belt-type respirometer on their upper torso. Each participant was seated and was breathing normally, and the measurements were acquired using two radar systems that were installed with different orientations and at different positions. Note that the belt-type respirometers were only used to evaluate the accuracy of the radar-based measurements. 

Both radar systems are frequency-modulated continuous wave (FMCW) radar systems with a center frequency of 79 GHz, a center wavelength of $\lambda=3.8$ mm, and an occupied bandwidth of 3.9 GHz. The beamwidths of the individual radar array elements are $\pm 4^\circ$ and $\pm 35^\circ$ in the E- and H-planes, respectively. The radar array is composed of a multiple-input multiple-output (MIMO) array that contains three transmitting and four receiving elements, with spacings of 7.6 mm ($2\lambda$) between the transmitting elements and spacings of 1.9\ mm ($\lambda/2$) between the receiving elements. In the experimental setting, the MIMO array can be approximated using a virtual linear array. Specifically, the array used in this study can be approximated using a 12-element virtual linear array with element spacings of $\lambda/2$. The slow-time sampling intervals were 100 ms. 

\begin{figure}[bt]
    \begin{center}
      \includegraphics[width=0.9\linewidth]{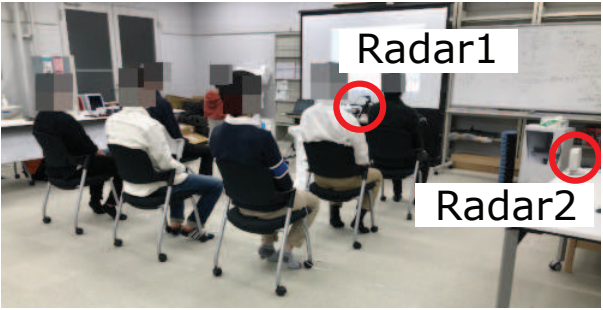}      
      \caption{Measurement setup with the seven seated participants shown with belt-type respirometer devices on their upper torsos in scenario 1.}
        \label{fig2}
    \end{center}
\end{figure}

\begin{figure}[bt]
    \begin{center}
      \includegraphics[width=0.7\linewidth]{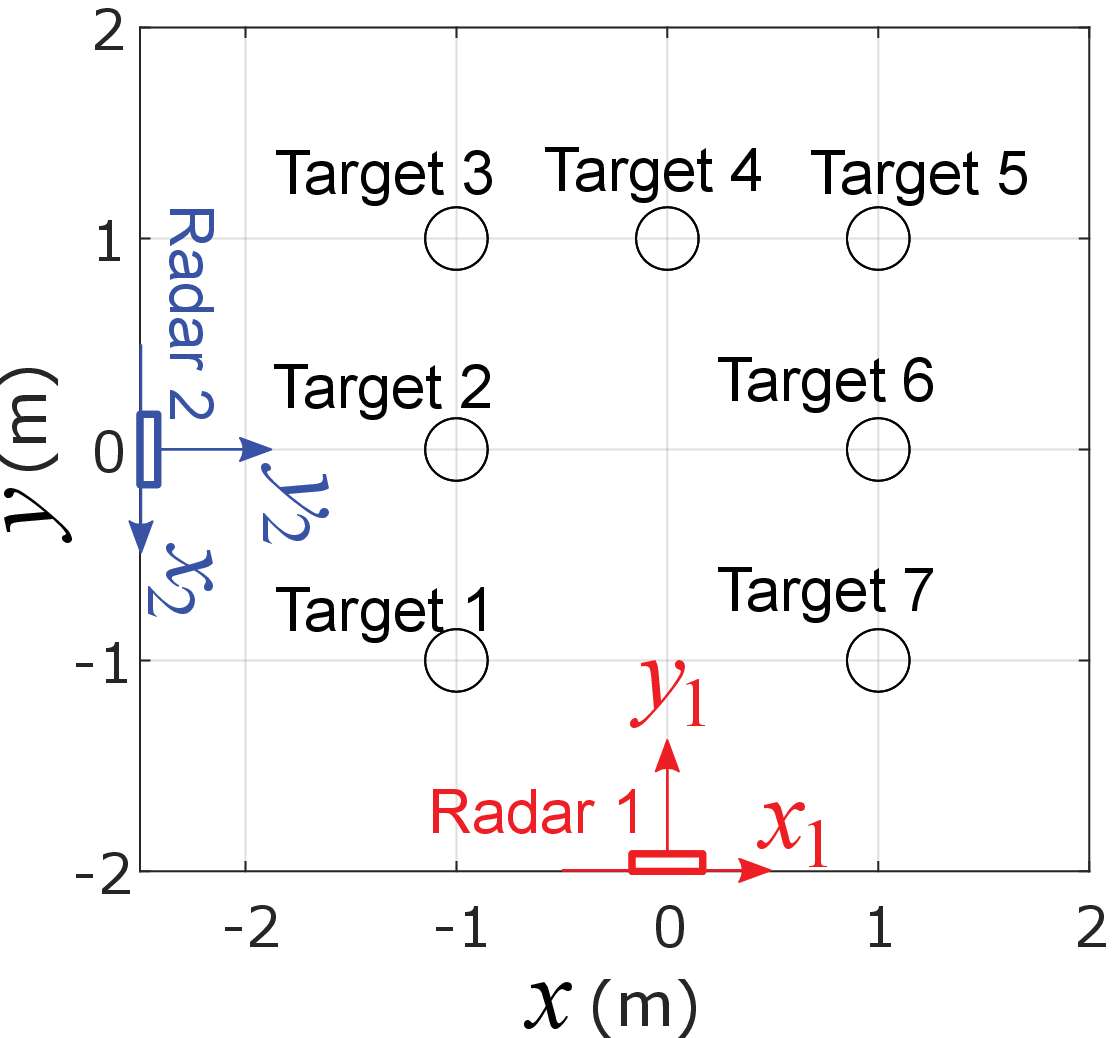}
      \caption{Actual layout of the radar systems and the participants in scenario 1.}
        \label{fig3}
    \end{center}
\end{figure}

\begin{figure}[bt]
  \begin{center}
    \includegraphics[width=0.7\linewidth]{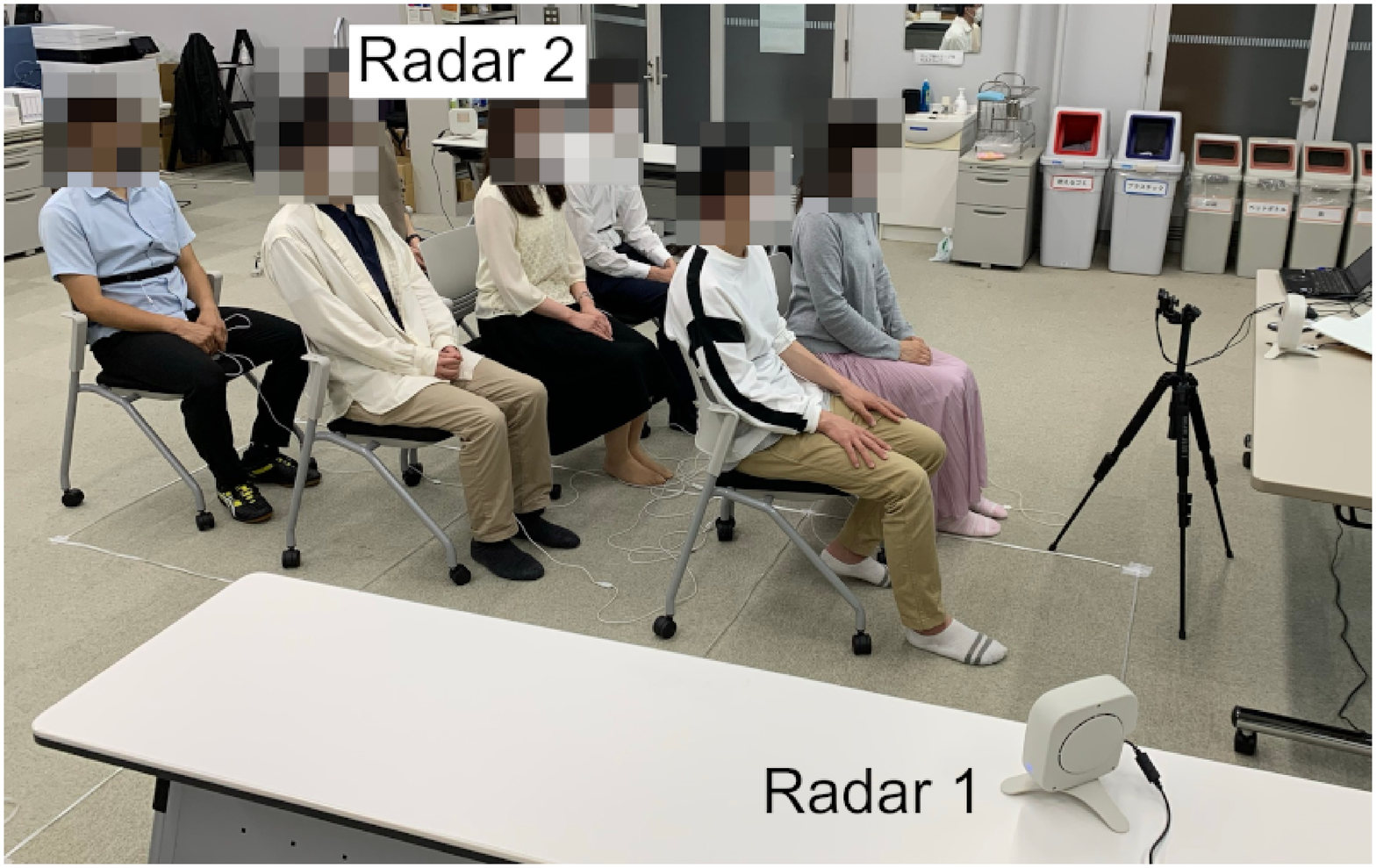}      
    \caption{Measurement setup with the seven seated participants shown with belt-type respirometer devices on their upper torsos in scenario 2.}
      \label{fig2b}
  \end{center}
\end{figure}

\begin{figure}[bt]
    \begin{center}
      \includegraphics[width=0.65\linewidth]{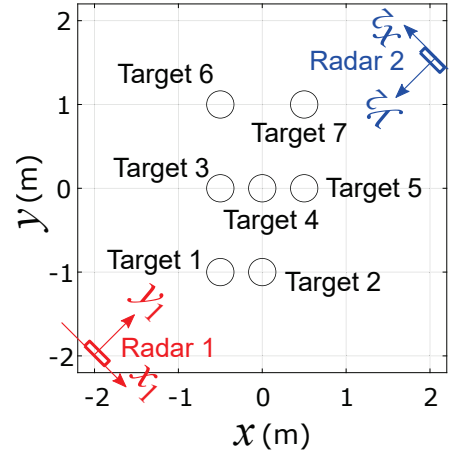}
      \caption{Actual layout of the radar systems and the participants in scenario 2.}
        \label{fig3b}
    \end{center}
\end{figure}

We apply the proposed method to the radar data that were measured with the seven participants in two scenarios. We set different layouts for both the human targets and the radar systems in these scenarios. In scenario 1, as shown in Fig.~\ref{fig2}, the seven participants were seated in a U-shaped arrangement with participant spacing of 1 m. We then used a pair of array radar systems to measure the participants as illustrated in Fig.~\ref{fig3}. In scenario 2, the seven participants were seated in a different layout, as shown in Fig.~\ref{fig2b}, and the radar systems were also placed at different positions with different orientations, as illustrated in Fig.\ref{fig3b}. The measurement time was set at 120 s and the participants were instructed to remain still and breathe normally during the measurements. To evaluate the accuracy of the respiration measurements, belt-type contact respirometers were used simultaneously with the radar measurements. 

\subsection{Application of the Proposed Data Fusion Method to the Measured Radar Data}
\subsubsection{Application in Scenario 1}
\begin{figure}[bt]
  \begin{center}
    \begin{minipage}[c]{0.4\linewidth}      
      \includegraphics[height=4cm]{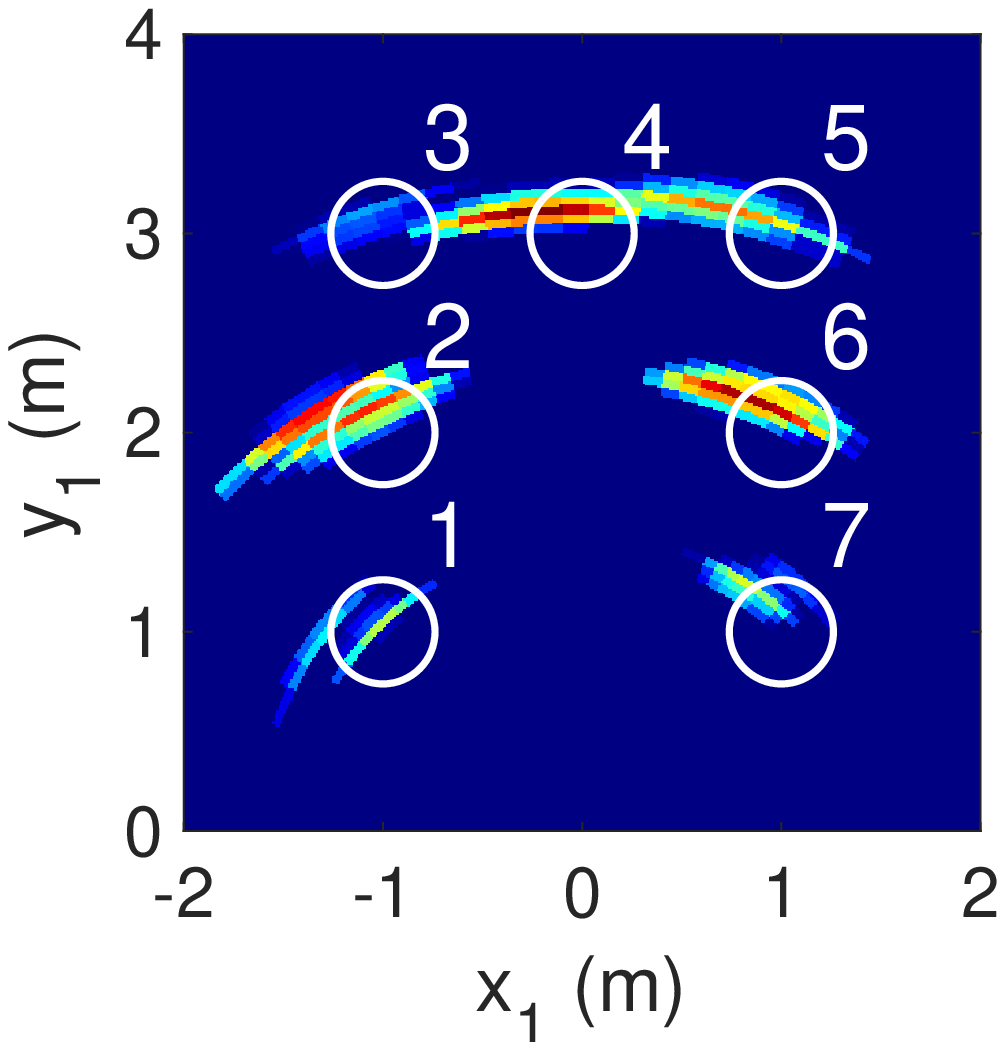}
    \end{minipage}
      \hspace{0.05\linewidth}    
    \begin{minipage}[c]{0.5\linewidth}
      \includegraphics[height=4cm]{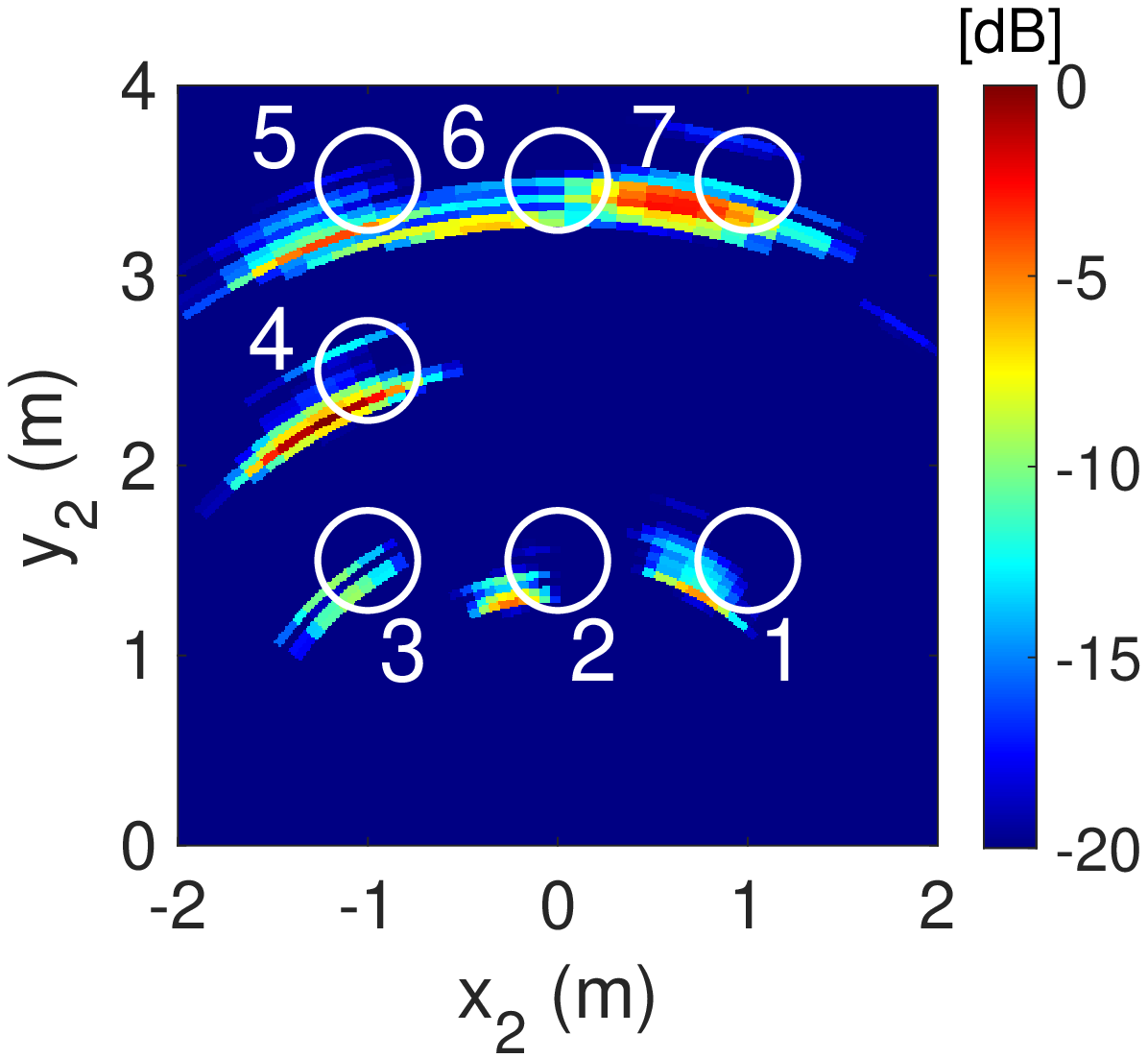}
    \end{minipage}
    \caption{Examples of radar images $|I_1(t,\vct{r}_1)|^2$ (left panel) and $|I_2(t,\vct{r}_2)|^2$ (right panel) at $t=30$ s in scenario 1.}
          \label{fig4}
  \end{center}
\end{figure}

\begin{figure}[bt]
  \begin{center}
      \begin{minipage}[c]{0.45\linewidth}
          \includegraphics[height=4cm]{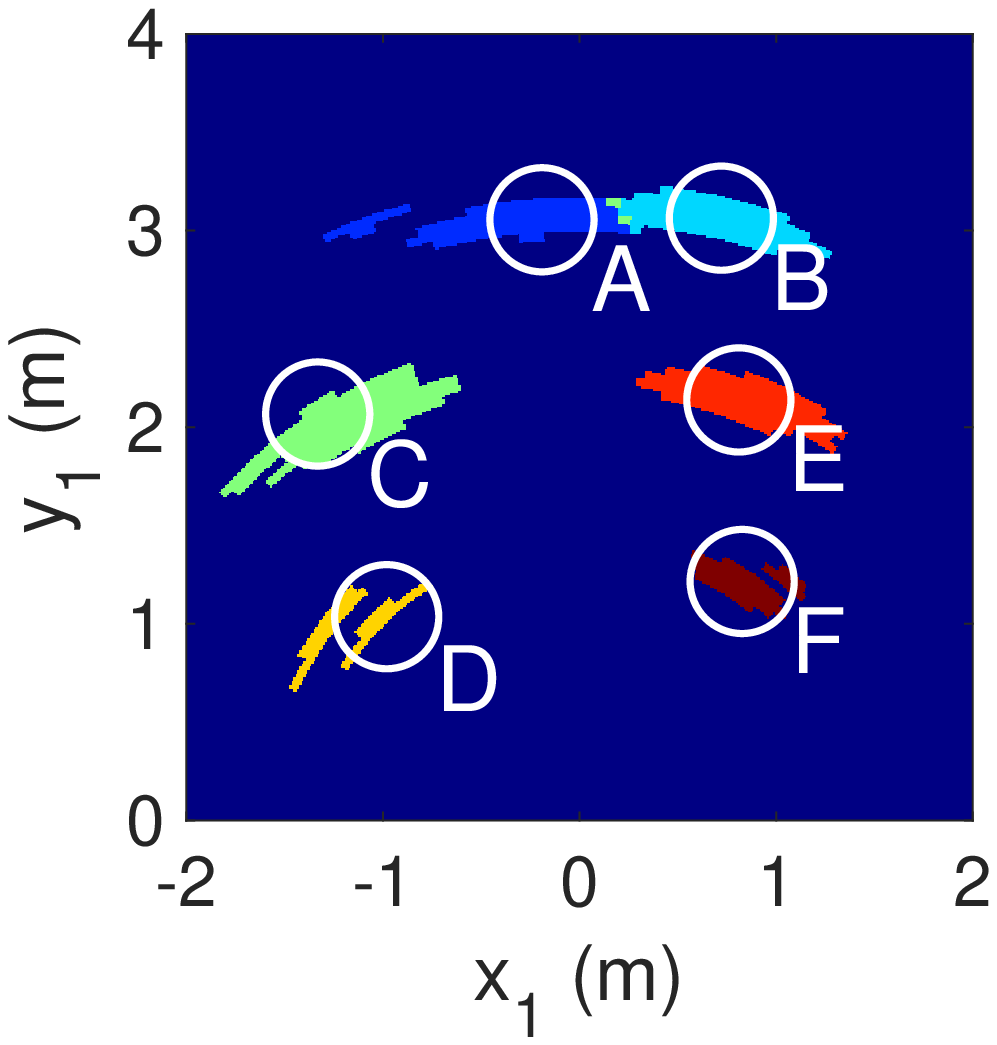}
      \end{minipage}
      \hspace{0.05\linewidth}
      \begin{minipage}[c]{0.45\linewidth}
          \includegraphics[height=4cm]{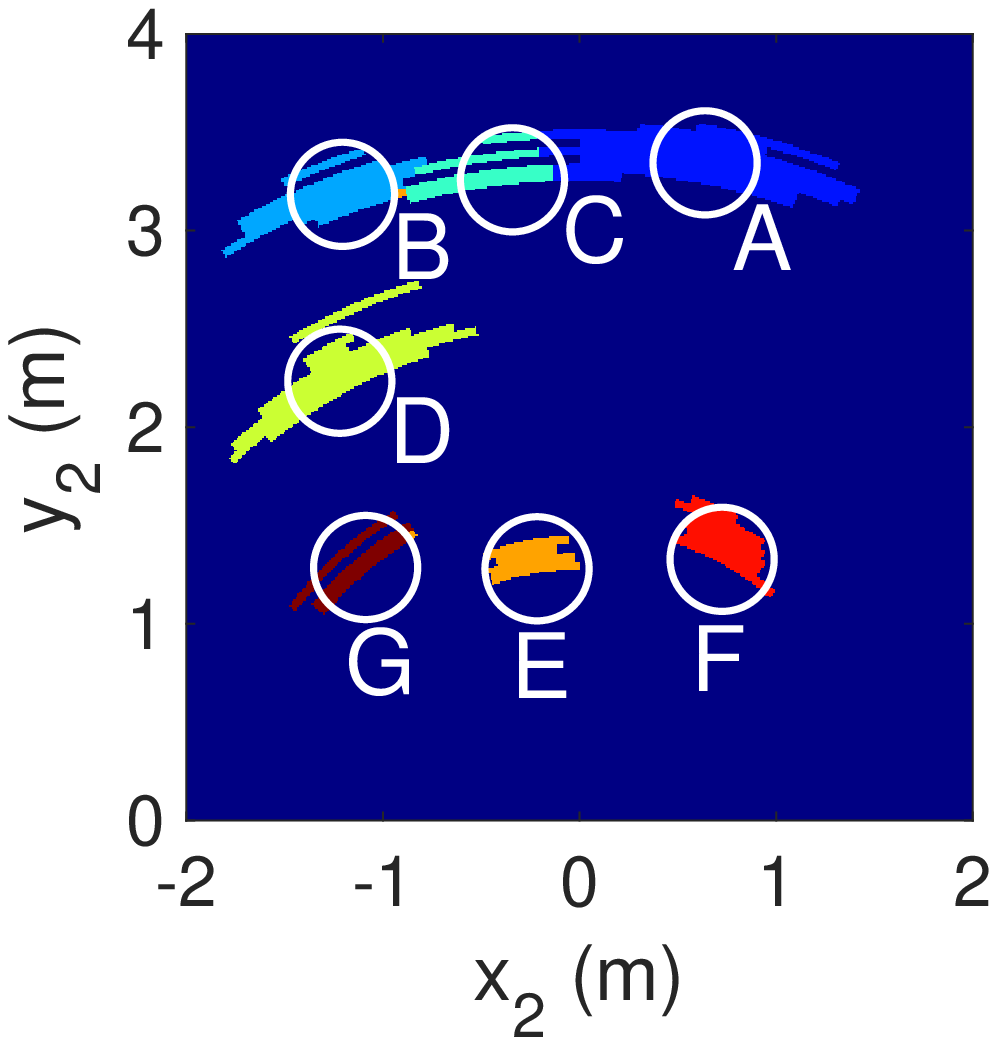}
      \end{minipage}
    \caption{Radar cluster images $|C_1(t,\vct{r}_1)|^2$ (left panel) and $|C_2(t,\vct{r}_2)|^2$ (right panel) at $t=30$ s in scenario 1. The different colors represent different clusters.}
    \label{fig5}
  \end{center}
\end{figure}

First, we apply the proposed method to the data acquired in scenario 1. The left and right panels of Fig.~\ref{fig4} show radar images $|I_1(t,\vct{r}_1)|^2$ and $|I_2(t,\vct{r}_2)|^2$ at $t=30$ s. In the figure, the actual target positions are indicated by white circles with the target numbers. Next, a point cloud is generated from these images and the respiratory-space clustering algorithm \cite{respiration21} is then applied to generate the radar cluster images $|C_1(t,\vct{r}_1)|^2$ and $|C_2(t,\vct{r}_2)|^2$ shown in Fig.~\ref{fig5}. In the figure, an alphabetical label is provided for each cluster. In image $|I_1(t,\vct{r}_1)|^2$ in Fig.~\ref{fig4}, the echo from target 3 is barely visible because of shadowing; as a result, in Fig.~\ref{fig5}, target 3 is not classified as a cluster. The resulting number of clusters was thus estimated erroneously to be six rather than seven, which is equivalent to the estimated number of human targets. This type of shadowing problem can occur particularly when multiple people are densely spaced within the measurement scene. To mitigate the shadowing problem, this study uses multiple radar systems rather than a single system.

Next, the proposed method is used to calculate the correlation coefficient $\rho_\mathrm{C}$ for each pair of displacement waveforms (see Table \ref{tab1}). The proposed method finds two pairs of points with the largest and second largest values of $\rho_\mathrm{C}$. In Table \ref{tab1}, we can find the largest correlation coefficient for the pairing of target E (radar 1) and target C (radar 2). We can also find the second largest correlation coefficient for the pairing of target C (radar 1) and target E (radar 2). These pairs $\mathrm{(E,C),(C,E)}$ are processed via the Procrustes analysis for $N=2$ and the rigid transformation parameters $x=-2.52\ \mathrm{m}$,\ $y=1.90\ \mathrm{m}$, and $\theta=-1.58\ \mathrm{rad}$ are obtained.

\begin{figure}[bt]
  \centering
      \begin{minipage}[c]{0.45\linewidth}
          \includegraphics[height=4cm]{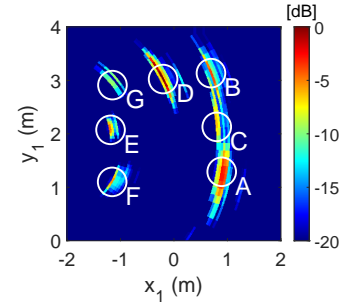}
      \end{minipage}
      \caption{Radar image $|\hat{I}_2(t,\vct{r}_1)|^2$ after the optimized rigid transformation is performed in scenario 1.}
    \label{fig6} 
 \end{figure}

Using these parameters, the radar image $|I_2(t,\vct{r}_2)|^2$ is then transformed into $|\hat{I}_2(t,\vct{r}_1)|^2$. Using the transformation parameters, the proposed method associates the six cluster pairs $\mathrm{(A,D)}$, $\mathrm{(B,B)}$, $\mathrm{(C,E)}$, $\mathrm{(D,F)}$, $\mathrm{(E,C)}$, and $\mathrm{(F,A)}$ from the seven human targets, where the distance threshold was set at $d_\mathrm{th}=0.5$ m. Note that cluster G from the second radar system was not associated in this process. The associated pairs are then processed via the Procrustes analysis and the transformation parameters $x=-2.46\ \mathrm{m}$,\ $y=1.86\ \mathrm{m}$, and $\theta=-1.57\ \mathrm{rad}$ are updated, as illustrated in Fig.~\ref{fig6}.

 \begin{table}[bt]
   \begin{center}
     \caption{Correlation Coefficients $\rho$ for the Displacement Waveforms in Scenario 1}
     \label{tab1}
     \begin{tabular}{|c|c||c|c|c|c|c|c|c|} \hline
       \multicolumn{2}{|c||}{}&\multicolumn{7}{c|}{Target index of radar 2} \\\cline{3-9}
       \multicolumn{2}{|c||}{}&A&B&C&D&E&F&G \\ \hline \hline 
               & A& 0.13& 0.63& 0.02& 0.15& 0.01& 0.17& 0.07 \\ \cline{2-9}
               & B& 0.19& 0.60& 0.10& 0.52& 0.34& 0.02& 0.02 \\ \cline{2-9}
       Target  & C& 0.20& 0.05& 0.04& 0.03& 0.81& 0.01& 0.01 \\ \cline{2-9}
      index of & D& 0.24& 0.08& 0.36& 0.06& 0.14& 0.36& 0.23 \\ \cline{2-9}
       radar 1 & E& 0.18& 0.07& 0.89& 0.04& 0.15& 0.13& 0.80 \\ \cline{2-9}
               & F& 0.04& 0.27& 0.56& 0.29& 0.04& 0.20& 0.51 \\ \hline
     \end{tabular}
   \end{center}
 \end{table}

 \begin{table}[tb]
  \begin{center}
    \caption{Target Detection Rates in Scenario 1}
    \label{tab2}
    \begin{tabular}{|c||c|c|c|c|c|c|c|}\hline
      &\multicolumn{7}{c|}{Participant number} \\ \cline{2-8}
      &1&2&3&4&5&6&7 \\ \hline \hline 
      Radar 1  (\%)& 100 & 100&   0 & 100& 100& 100& 100\\ \hline
      Radar 2  (\%)& 100 & 100& 100 & 100& 100&  69& 100\\ \hline
      Proposed (\%)& 100 & 100& 100 & 100& 100& 100& 100\\ \hline
    \end{tabular}
  \end{center}
\end{table}

For the associated cluster pair $(n_1,n_2)$, the respiratory displacement waveforms $\angle I_1(t,\vct{r}_1^{(n_1)})$ and $\angle I_2(t,\vct{r}_2^{(n_2)})$ are then obtained. The detection rate for 120 s for each target is shown in Table \ref{tab2}. The table shows that human target 3 was not detected by one of the radar systems because of shadowing. All the human targets, with the exception of target 3, were detected by both radar systems, which gives us selection options; we can thus select a more accurate radar system for each target. In the proposed method, the fourth-order statistical parameter $\kappa$ is calculated for both displacement waveforms, from which the waveform with the larger $\kappa$ value is selected every 30 s. Table \ref{tab5} shows the errors in estimation of the respirations per minute (rpm) obtained when the two radar systems are combined. The error averaged over the seven participants is 0.33 rpm for the proposed process when combining the two radar systems. These results indicate that the proposed method using the multiradar system is effective in performing accurate respiration measurements of multiple people simultaneously.

 \begin{table}[bt]
   \begin{center}
     \caption{Correlation Coefficients $\rho$ for the Displacement Waveforms in Scenario 2}
     \label{tab3}
     \begin{tabular}{|c|c||c|c|c|c|c|c|} \hline
       \multicolumn{2}{|c||}{}&\multicolumn{6}{c|}{Target index of radar 2} \\\cline{3-8}
       \multicolumn{2}{|c||}{}&A&B&C&D&E&F \\ \hline \hline 
                 & A & 0.27& 0.38& 0.14& 0.22& 0.03& 0.38\\ \cline{2-8}
        Target   & B & 0.01& 0.14& 0.36& 0.03& 0.02& 0.20\\ \cline{2-8}
        index of & C & 0.25& 0.42& 0.18& 0.38& 0.11& 0.44\\ \cline{2-8}
        radar 1  & D & 0.31& 0.04& 0.29& 0.21& 0.35& 0.15\\ \cline{2-8}
                 & E & 0.09& 0.07& 0.08& 0.29& 0.07& 0.17\\ \hline
     \end{tabular}
   \end{center}
 \end{table}

\begin{table}[tb]
  \begin{center}
    \caption{Target Detection Rates in Scenario 2}
    \label{tab4}
    \begin{tabular}{|c||c|c|c|c|c|c|c|}\hline
      &\multicolumn{7}{c|}{Participant number} \\ \cline{2-8}
      &1&2&3&4&5&6&7 \\ \hline \hline 
      Radar 1  (\%)& 100 & 100& 100 & 100&   0&  58&  72\\ \hline
      Radar 2  (\%)&   0 & 100& 100 & 100& 100& 100& 100\\ \hline
      Proposed (\%)& 100 & 100& 100 & 100& 100& 100& 100\\ \hline
    \end{tabular}
  \end{center}
\end{table}

\subsubsection{Application in Scenario 2}
\begin{figure}[bt]
  \begin{center}
    \begin{minipage}[c]{0.4\linewidth}      
      \includegraphics[height=4cm]{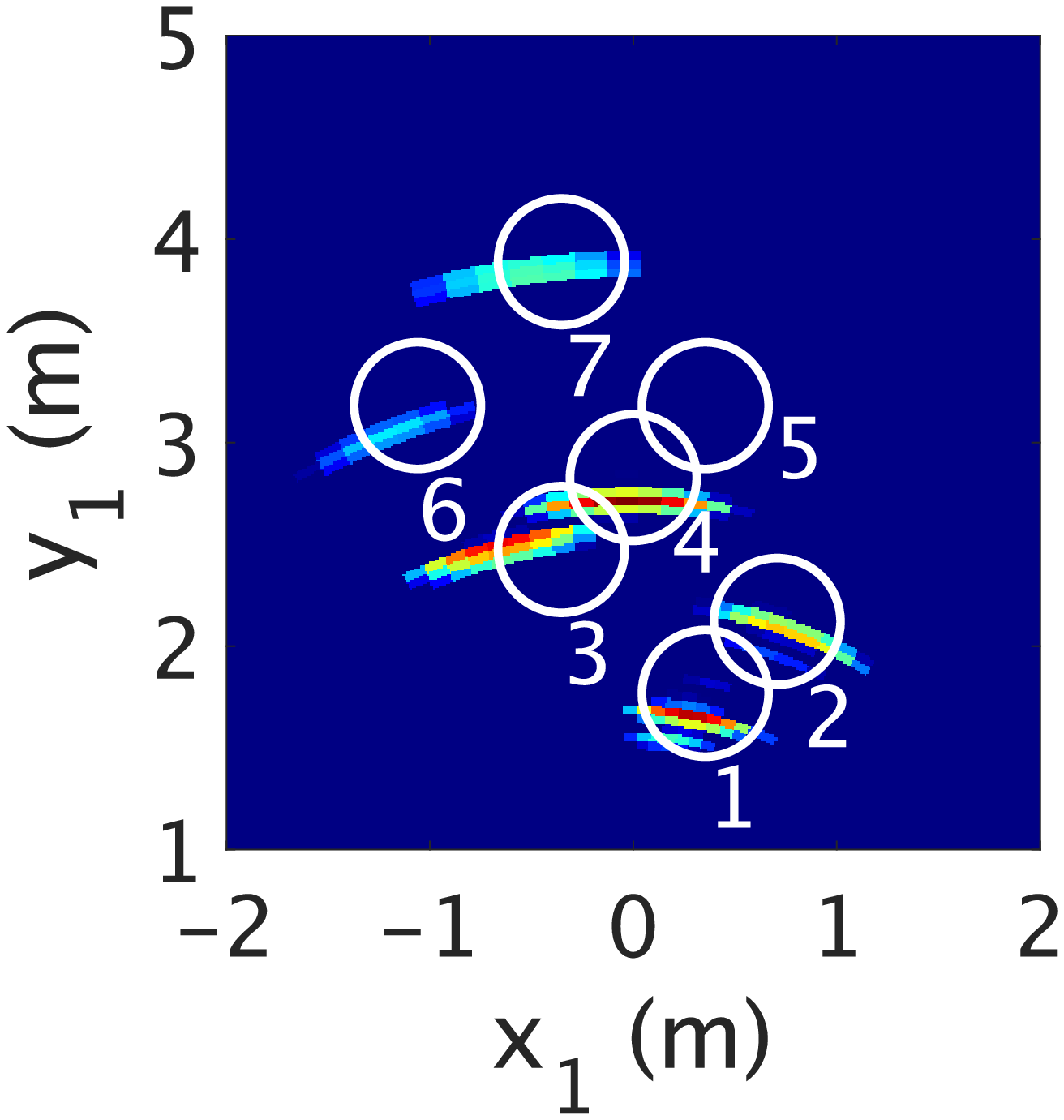}
    \end{minipage}
      \hspace{0.05\linewidth}    
    \begin{minipage}[c]{0.5\linewidth}
      \includegraphics[height=4cm]{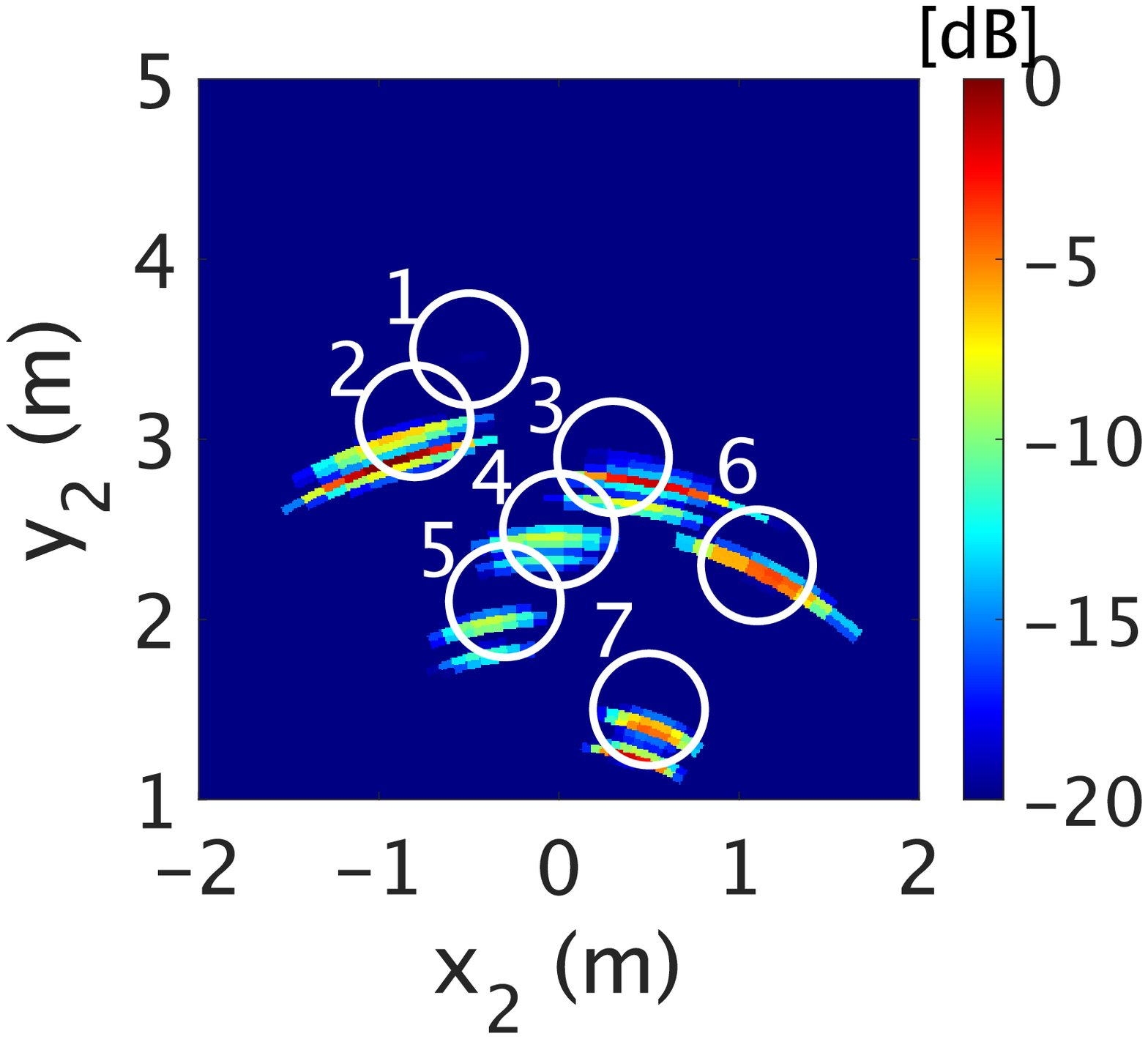}
    \end{minipage}
    \caption{Examples of radar images $|I_1(t,\vct{r}_1)|^2$ (left panel) and $|I_2(t,\vct{r}_2)|^2$ (right panel) at $t=30$ s in scenario 2.}
          \label{fig4b}
  \end{center}
\end{figure}

\begin{figure}[bt]
  \begin{center}
      \begin{minipage}[c]{0.4\linewidth}
          \includegraphics[height=4cm]{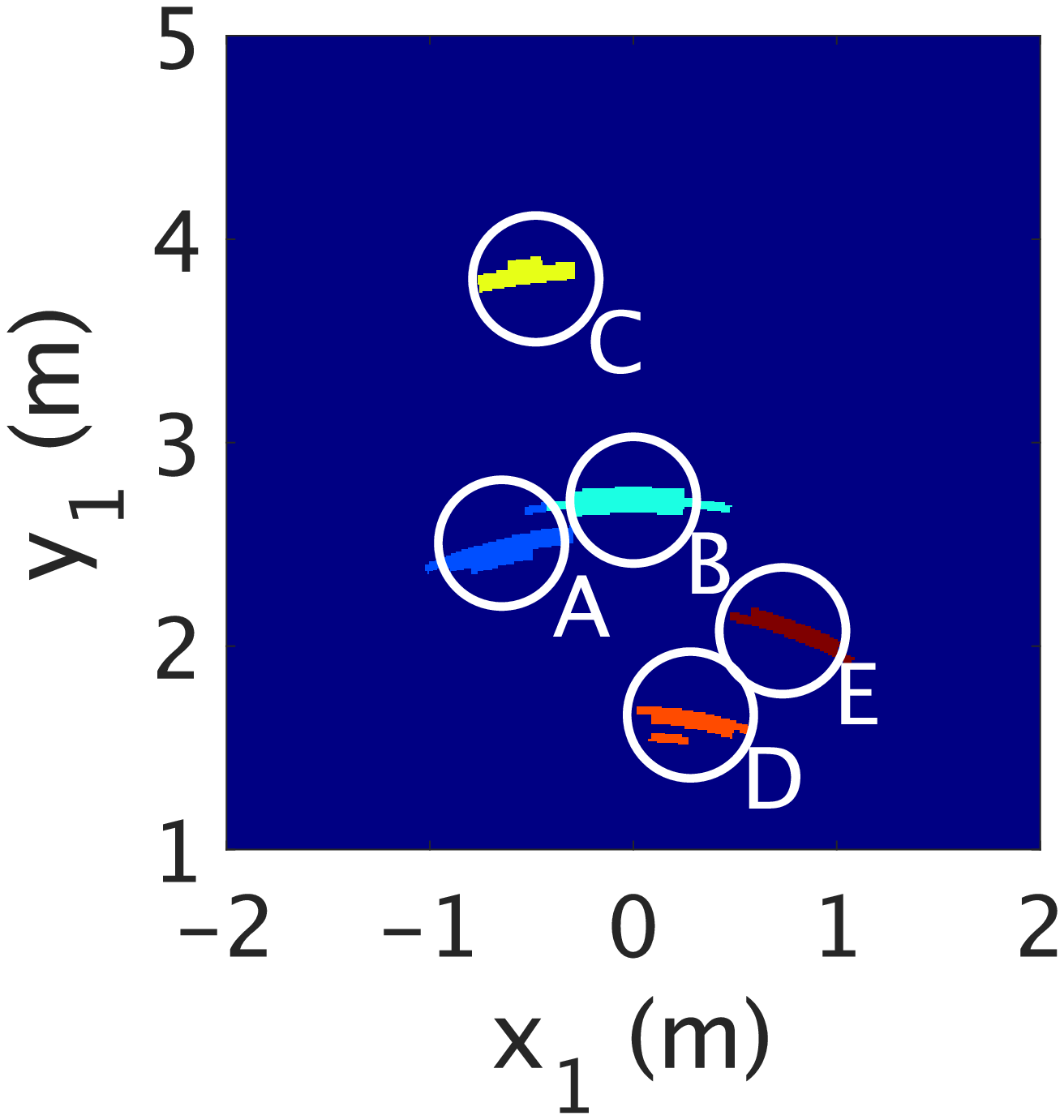}
      \end{minipage}
      \hspace{0.05\linewidth}
      \begin{minipage}[c]{0.4\linewidth}
          \includegraphics[height=4cm]{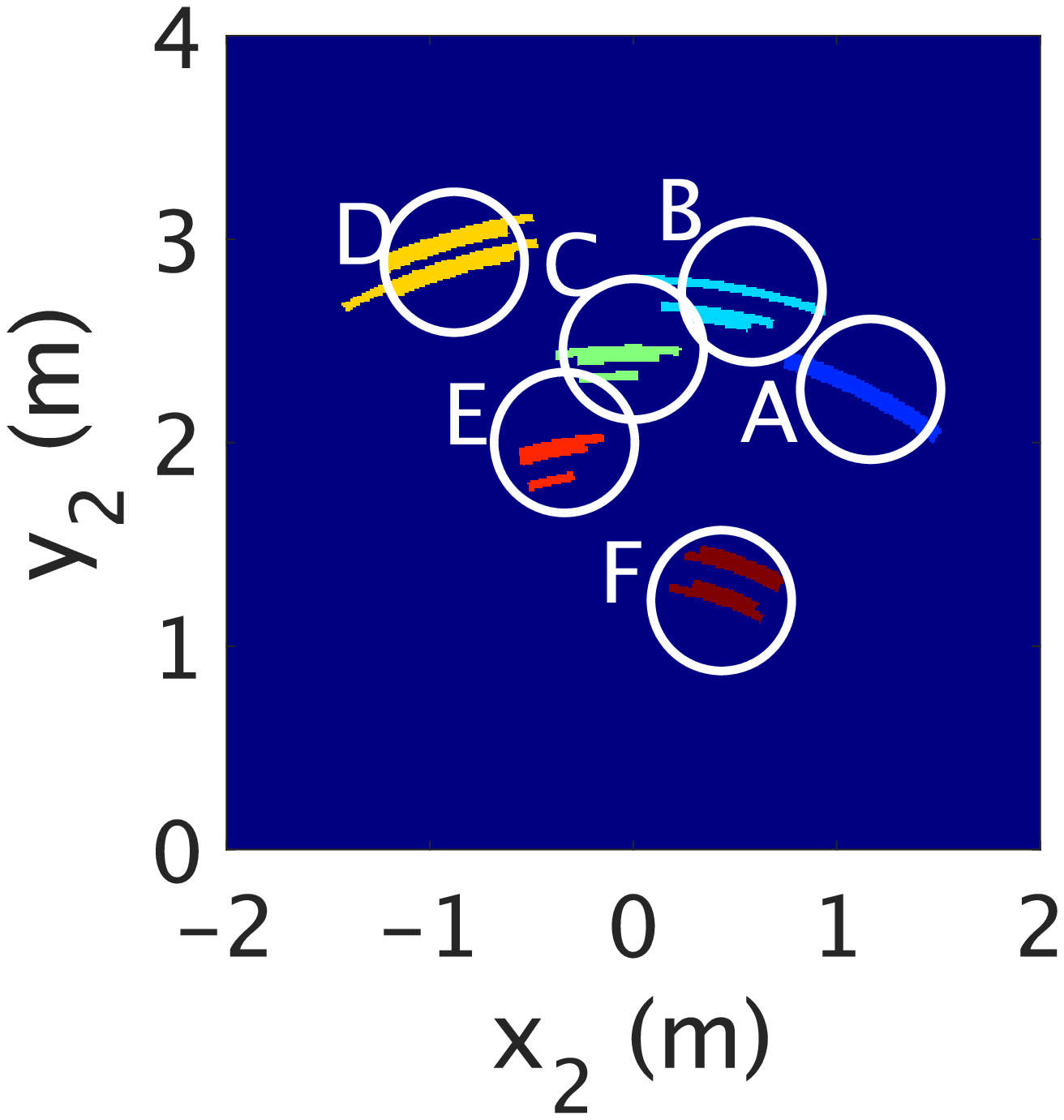}
      \end{minipage}
      \hspace{0.05\linewidth}
    \caption{Radar cluster images $|C_1(t,\vct{r}_1)|^2$ (left panel) and $|C_2(t,\vct{r}_2)|^2$ (right panel) at $t=30$ s in scenario 2. The different colors represent different clusters.}
    \label{fig5b}
  \end{center}
\end{figure}

We also applied the proposed method to the radar data acquired in scenario 2. The layout of the target people in this case is shown in Fig.~\ref{fig3b}. Using radar systems 1 and 2, the radar images $|I_1(t,\vct{r}_1)|^2$ and $|I_2(t,\vct{r}_1)|^2$ were generated as shown in Fig.~\ref{fig4b}. We can then generate the radar cluster images $|C_1(t,\vct{r}_1)|^2$ and $|C_2(t,\vct{r}_2)|^2$ shown in Fig.~\ref{fig5b}. In Fig.~\ref{fig5b}, no clusters are generated for targets 5 and 6 because the echoes from targets 5 and 6 are weak in image $|I_1(t,\vct{r}_1)|^2$, as shown in Fig.~\ref{fig4b}. Similarly, no cluster is generated for target 1 because the echo is weak in image $|I_2(t,\vct{r}_1)|^2$, as shown in Fig.~\ref{fig4b}.

The proposed method is then used to calculate the correlation coefficient $\rho_\mathrm{C}$ for each of the pairs of displacement waveforms (Table \ref{tab3}). In Table \ref{tab3}, we find that the largest correlation coefficient is that for the pairing of target C (radar 1) and target F (radar 2), and also find that the second largest correlation coefficient is that for the pairing of target A (radar 1) and target B (radar 2). These pairs, denoted by $\mathrm{(C,F),(A,B)}$, are processed via the Procrustes analysis for $N=2$ and the rigid transformation parameters $x=-0.17\ \mathrm{m}$,\ $y=5.11\ \mathrm{m}$, and $\theta=-3.07\ \mathrm{rad}$ are obtained.

\begin{figure}[bt]
  \centering
      \begin{minipage}[c]{0.45\linewidth}
          \includegraphics[height=4cm]{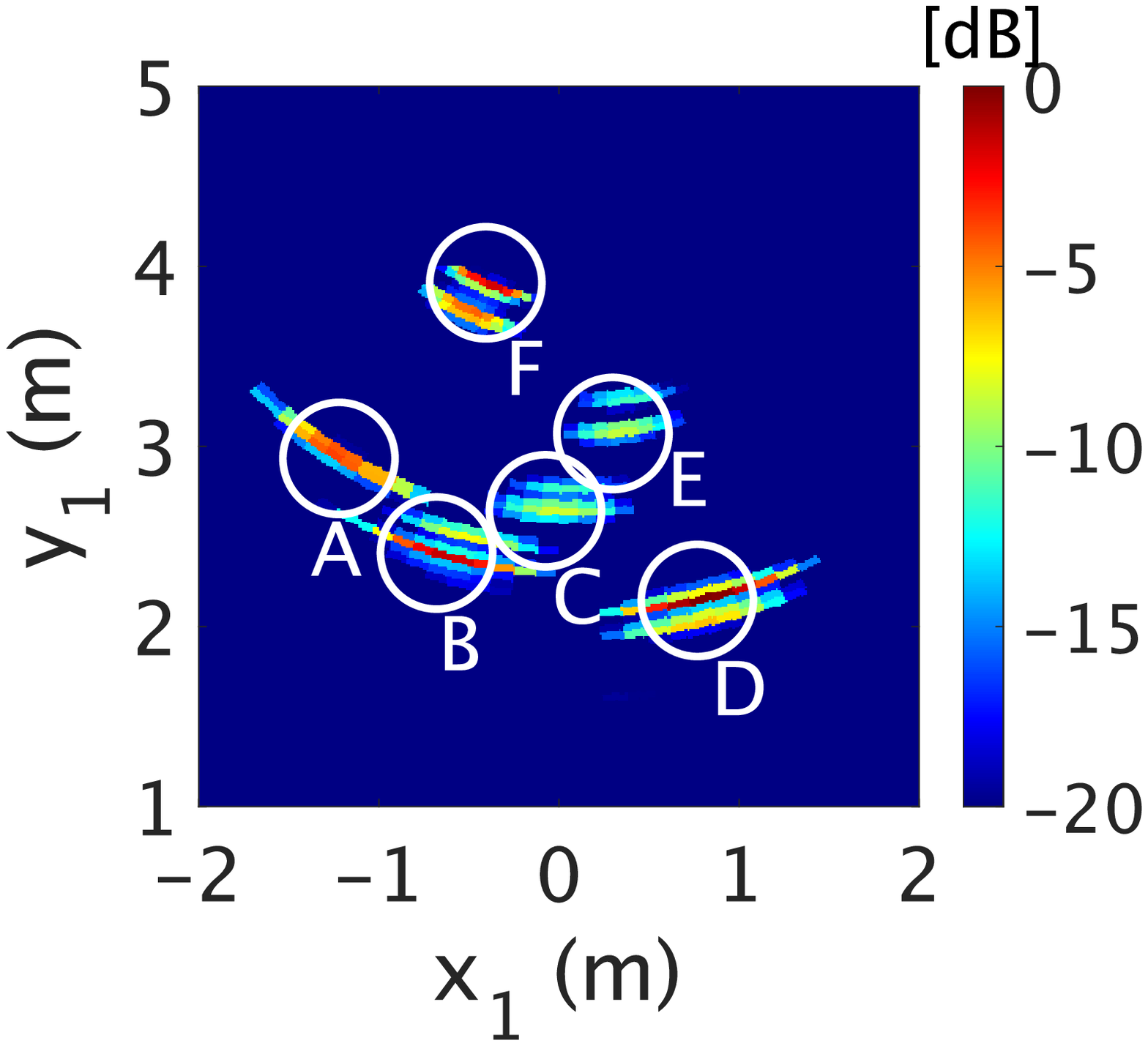}
      \end{minipage}
      \caption{Radar image $|\hat{I}_2(t,\vct{r}_1)|^2$ after the optimized rigid transformation is performed in scenario 2.}
    \label{fig6b} 
 \end{figure}

Then, using these parameters, the radar image $|I_2(t,\vct{r}_2)|^2$ is transformed into $|\hat{I}_2(t,\vct{r}_1)|^2$. Using the transformation parameters, the proposed method associates the four cluster pairs $\mathrm{(A,B)}$, $\mathrm{(B,C)}$, $\mathrm{(C,F)}$, and $\mathrm{(E,D)}$ from the clusters, where the distance threshold was set at $d_\mathrm{th}=0.5$ m. Note that cluster D from the first radar system and clusters A and E from the second radar system were not associated in this process. Using the associated pairs, the transformation parameters $x=0.12\ \mathrm{m}$,\ $y=5.09\ \mathrm{m}$, and $\theta=3.06\ \mathrm{rad}$ are updated, as illustrated in Fig.~\ref{fig6b}.

For the associated cluster pair $(n_1,n_2)$, the respiratory displacement waveforms $\angle I_1(t,\vct{r}_1^{(n_1)})$ and $\angle I_2(t,\vct{r}_2^{(n_2)})$ are obtained. The respiration detection rates for each target when using the proposed method are shown in Table \ref{tab4}, which shows that human targets 1 and 5 were only detected by one of the two radar systems because of shadowing, whereas the other targets were detected by both radar systems. As in scenario 1, the fourth-order statistical parameter $\kappa$ is calculated for each of the estimated displacement waveforms, from which the waveform with the larger value of $\kappa$ is selected every 30 s. Table \ref{tab5} shows the errors in respiration rate estimation when using the proposed data fusion technique for these radar systems. The average error for the seven participants was 1.24 rpm when using the proposed method to combine the radar data.

\begin{table}[tb]
  \begin{center}
    \caption{Errors in RPM Estimation when Using the Proposed Method}
    \label{tab5}
    \begin{tabular}{|c||c|c|c|c|c|c|c|}\hline
      &\multicolumn{7}{c|}{Participant number} \\ \cline{2-8}
      &1&2&3&4&5&6&7 \\ \hline \hline 
      Scenario 1 (rpm)& 0.43 & 0.02& 0.06& 0.88& 0.72& 0.05& 0.15\\ \hline
      Scenario 2 (rpm)& 0.40 & 1.79& 1.75& 1.97& 0.04& 0.38& 2.35\\ \hline
    \end{tabular}
  \end{center}
\end{table}
 
\section{Conclusion}
In this study, we have proposed a novel method for a multiradar system that can measure the positions and respiration characteristics of multiple people. By introducing multiple radar systems rather than a single system, the shadowing problem can be mitigated, which means that the respiration rates of multiple people can be measured even when these people are located closely together. In addition, when using multiple radar systems, we can measure each person from multiple directions, which improves the accuracy of the respiration measurements. The effectiveness of the proposed method was verified experimentally in two scenarios using a pair of radar systems and seven participants.

In both scenarios, the proposed method was demonstrated to be able to detect all participants by combining the data acquired from the two radar systems, whereas not all participants could be detected when using only one of the two radar systems because of shadowing. The average respiratory rate estimation error for the seven participants in the two scenarios was 0.79 rpm, which is sufficiently small when compared with the typical respiration rate range for adults (12-20 rpm). Use of the multiradar system with the proposed method has been demonstrated to be promising for use in accurate noncontact monitoring of the respiration of multiple people.

\section*{Acknowledgment}
We thank Dr. Hirofumi Taki and Dr. Shigeaki Okumura of MaRI Co., Ltd. for their help with this study.

\section*{Ethics Declarations}
This study was approved by the Ethics Committee of the Graduate School of Engineering, Kyoto University (permit no.~201916). Informed consent was obtained from all participants in the study.

\newpage
\begin{IEEEbiography}[{\includegraphics[width=1in,height=1.25in,clip,keepaspectratio]{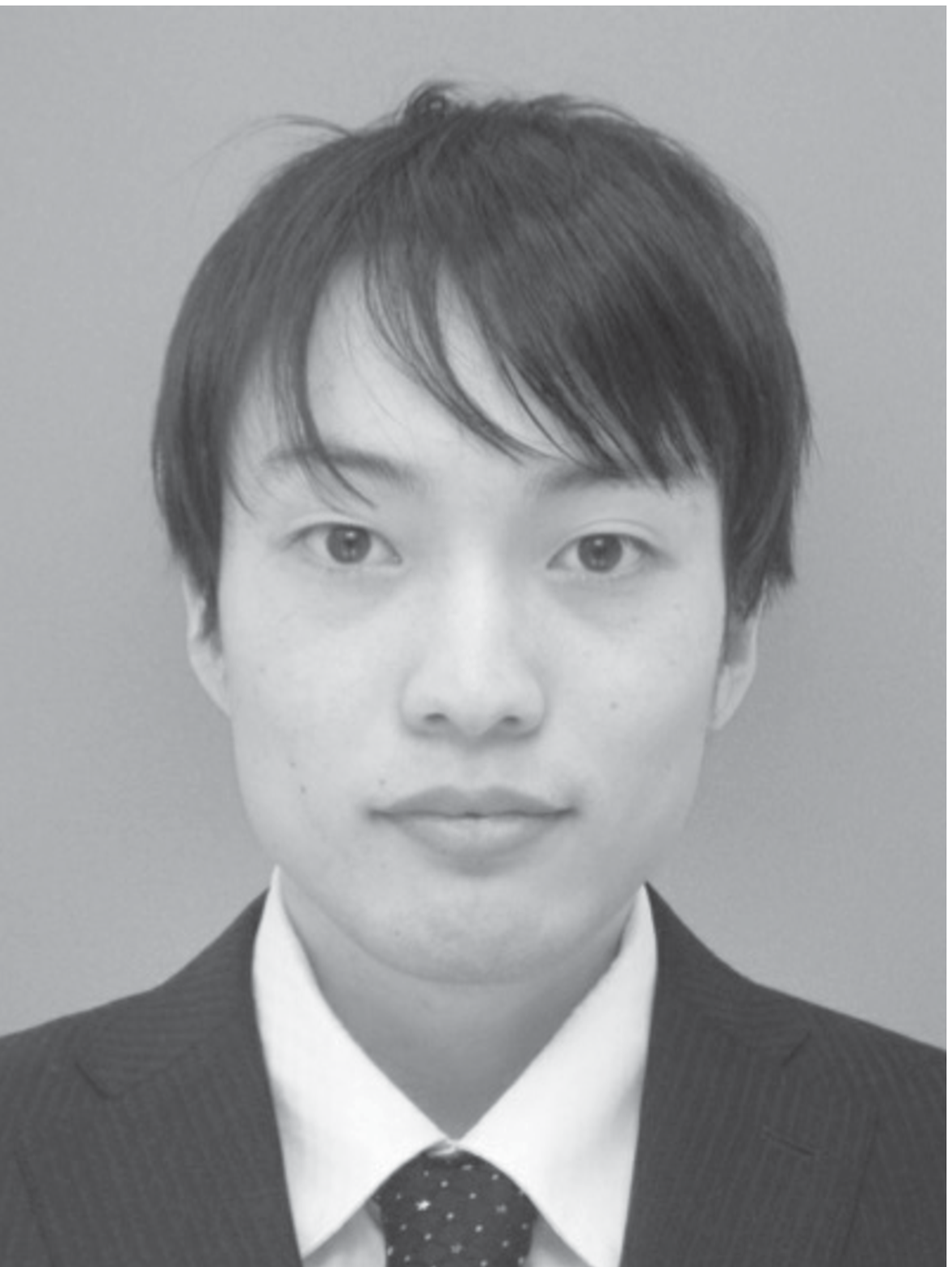}}]{Shunsuke Iwata} (STM'21)
  received a B.E.~degree in electrical and electronic engineering from Kyoto University, Kyoto, Japan, in 2021. He is currently working toward the M.E. degree in electrical engineering at the Graduate School of Engineering, Kyoto University. Mr. Iwata is a recipient of 2020 IEEE AP-S Kansai Joint Chapter Best Presentation Award. His research interests include multiradar measurement of respiration of multiple people.
\end{IEEEbiography}
\vskip 0pt plus -1fil

\begin{IEEEbiography}[{\includegraphics[width=1in,height=1.25in,clip,keepaspectratio]{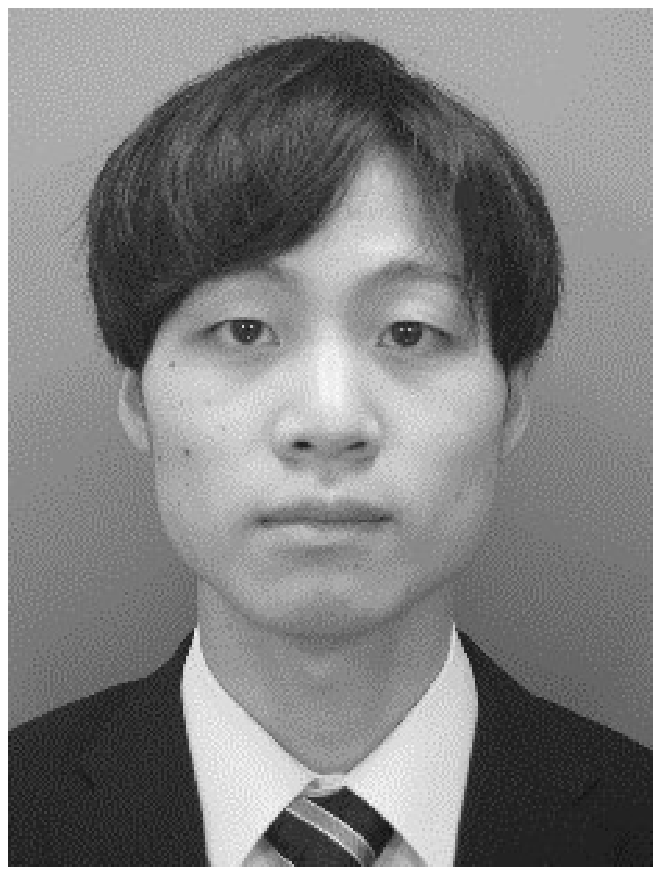}}]{Takato Koda}
 received a B.E.~degree in electrical and electronic engineering from Kyoto University, Kyoto, Japan, in 2020. He is currently working toward the M.E. degree in electrical engineering at the Graduate School of Engineering, Kyoto University. His research interests include radar imaging and clustering of multiple people using array radar systems.
\end{IEEEbiography}
\vskip 0pt plus -1fil

\begin{IEEEbiography}[{\includegraphics[width=1in,height=1.25in,clip,keepaspectratio]{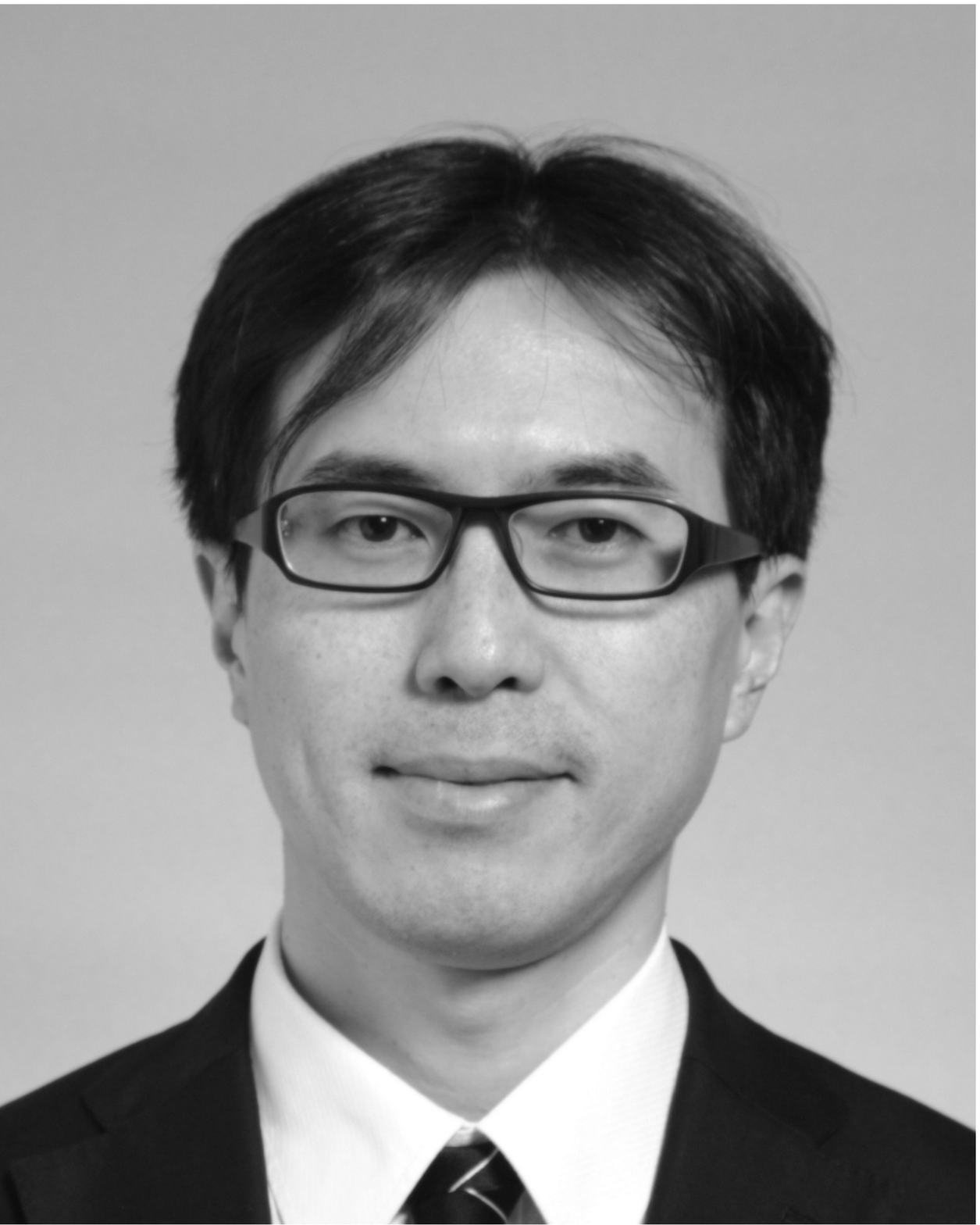}}]{Takuya Sakamoto} (M'04--SM'17)
received a B.E.~degree in electrical and electronic engineering from
Kyoto University, Kyoto, Japan, in 2000, and M.I.~and Ph.D.~degrees in
communications and computer engineering from the Graduate School of
Informatics, Kyoto University, in 2002 and 2005, respectively.

From 2006 to 2015, he was an Assistant Professor at the Graduate School
of Informatics, Kyoto University. From 2011 to 2013, he was also a
Visiting Researcher at Delft University of Technology, Delft, the
Netherlands. From 2015 to 2018, he was an Associate Professor at the
Graduate School of Engineering, University of Hyogo, Himeji, Japan. In
2017, he was also a Visiting Scholar at the University of Hawaii at
Manoa, Honolulu, HI, USA. Since 2018, he has been a PRESTO Researcher at
the Japan Science and Technology Agency, Kawaguchi, Japan. At present, he
is an Associate Professor at the Graduate School of Engineering, Kyoto
University. His current research interests are system theory, inverse
problems, radar signal processing, radar imaging, and wireless sensing
of vital signs.

Dr. Sakamoto was a recipient of the Best Paper Award from the
International Symposium on Antennas and Propagation (ISAP) in 2012, and
the Masao Horiba Award in 2016. In 2017, he was invited as a
semi-plenary speaker to the European Conference on Antennas and
Propagation (EuCAP) in Paris, France.
\end{IEEEbiography}

\end{document}